%% file: main.tex
\documentclass[conference]{IEEEtran}

\usepackage{cite}
\usepackage{amsmath,amssymb,amsfonts}
\usepackage{algorithmic}
\usepackage{graphicx}
\usepackage{textcomp}
\usepackage{xcolor}
\usepackage{algorithm2e}
\usepackage{fancyhdr}

\newcommand{\ve}[1]{\mathbf{#1}} 

\usepackage[nolist]{acronym}
\input{acronyms.tex}

\def\BibTeX{{\rm B\kern-.05em{\sc i\kern-.025em b}\kern-.08em
    T\kern-.1667em\lower.7ex\hbox{E}\kern-.125emX}}
\fancypagestyle{FirstPage}{
}
\begin{document}

\title{RNN-Based GNSS Positioning using Satellite Measurement Features and Pseudorange Residuals}

\author{\IEEEauthorblockN{Ibrahim Sbeity}
\IEEEauthorblockA{\textit{CEA-Leti} \\
\textit{Universit\'e Grenoble Alpes}\\
F-38000 Grenoble, France \\
\textit{ETIS UMR 8051} \\
\textit{CY Cergy Paris Universit\'e, ENSEA, CNRS}\\
F-95000 Cergy, France \\
Ibrahim.Sbeity@cea.fr}
\and
\IEEEauthorblockN{Christophe Villien}
\IEEEauthorblockA{\textit{CEA-Leti} \\
\textit{Universit\'e Grenoble Alpes}\\
F-38000 Grenoble, France \\
Christophe.Villien@cea.fr}
\and
\IEEEauthorblockN{Marwa Chafii}
\IEEEauthorblockA{\textit{Engineering Division} \\
\textit{New York University (NYU) Abu Dhabi}\\
Abu Dhabi, UAE \\
\textit{NYU WIRELESS} \\
\textit{NYU Tandon School of Engineering}\\
Brooklyn, NY 11201, USA \\
email address or ORCID}
\and
\IEEEauthorblockN{Christophe Combettes}
\IEEEauthorblockA{\textit{CEA-Leti} \\
\textit{Universit\'e Grenoble Alpes}\\
F-38000 Grenoble, France \\
Christophe.Combettes@cea.fr}
\and
\IEEEauthorblockN{Benoît Denis}
\IEEEauthorblockA{\textit{CEA-Leti} \\
\textit{Universit\'e Grenoble Alpes}\\
F-38000 Grenoble, France \\
Benoit.Denis@cea.fr}
\and
\IEEEauthorblockN{E. Veronica Belmega}
\IEEEauthorblockA{\textit{dept. name of organization (of Aff.)} \\
\textit{Univ. Gustave Eiffel, CNRS, LIGM}\\
F-77454 Marne-la-Valle\'e, France \\
\textit{ETIS UMR 8051} \\
\textit{CY Cergy Paris Universit\'e, ENSEA, CNRS}\\
F-95000 Cergy, France \\
email address or ORCID}
}

\author{\IEEEauthorblockN{Ibrahim Sbeity\IEEEauthorrefmark{1}\IEEEauthorrefmark{2},
Christophe Villien\IEEEauthorrefmark{1}, 
Benoît Denis\IEEEauthorrefmark{1}, and E. Veronica Belmega\IEEEauthorrefmark{3}\IEEEauthorrefmark{2}}\\
\IEEEauthorblockA{
\IEEEauthorrefmark{1}
CEA-Leti, Universit\'e Grenoble Alpes, 
F-38000 Grenoble, France \\
\IEEEauthorrefmark{2} ETIS UMR 8051, CY Cergy Paris Universit\'e, ENSEA, CNRS, F-95000, Cergy, France\\
\IEEEauthorrefmark{3} Univ. Gustave Eiffel, CNRS, LIGM, F-77454,  Marne-la-Vallée, France\\
Emails: ibrahim.sbeity@cea.fr, christophe.villien@cea.fr}
}

\maketitle
\begin{abstract}
\textcolor{black}{In the \ac{GNSS} context,} \textcolor{black}{the growing number of available satellites has lead to many challenges when} \textcolor{black}{it comes to choosing} \textcolor{black}{the most accurate} \textcolor{black}{pseudorange contributions,} \textcolor{black}{given the strong impact of} 
\textcolor{black}{biased measurements} \textcolor{black}{on positioning accuracy, particularly in single-epoch scenarios. This work leverages the potential of machine learning in predicting link-wise} \textcolor{black}{measurement} \textcolor{black}{quality factors} \textcolor{black}{and, hence, optimize measurement weighting}. \textcolor{black}{For this purpose, we use a customized matrix composed of heterogeneous features such as conditional pseudorange residuals and per-link satellite} \textcolor{black}{metrics} \textcolor{black}{(e.g., carrier-to-noise power density ratio and its} \textcolor{black}{empirical} \textcolor{black}{statistics, satellite elevation, carrier phase lock time). This matrix is then fed as an input to a \ac{RNN} (i.e., a \ac{LSTM} network). Our experimental results on real data, obtained from extensive field measurements, demonstrate the high potential of our proposed solution being able to outperform traditional measurements weighting and selection strategies from state-of-the-art.}

\end{abstract}

\thispagestyle{FirstPage}
\vspace{-2mm}
\section{Introduction}

Accurate outdoor positioning using \acf{GNSS} is a key enabler for many important applications, such as autonomous vehicles and unmanned aerial vehicles, tracking of blue force and first responders, seamless end-to-end logistics and supply chains optimization, large-scale crowd sensing, etc. To ensure reliable positioning, several \ac{SV} selection, \ac{FDE}, and weighting techniques, must be performed during the tracking phase of the navigation processor. 
However, in a single-epoch context, the problem becomes even more challenging, taking into consideration that there is no access to prior estimates. 

Despite these challenges, single-epoch positioning remains essential in several operating contexts. First, it can serve as an initial position estimate to initialize the navigation processor \cite{grewal2007global}. \textcolor{black}{Second, the loosely-coupled data fusion of GNSS with an} \textcolor{black}{\ac{INS}} \textcolor{black}{requires the GNSS outputs (i.e., position and velocity) to be independent. This is to prevent any time-correlated measurements from occurring at the input of the fusion engine. Therefore, single-epoch solutions are necessary. Finally, recently developed} \textcolor{black}{\ac{IoT}} \textcolor{black}{chips, such as Semtech's LoRa chip LR1110 \cite{simtechLoraManual}, utilize low-power receivers to capture a single snapshot of \ac{GNSS} measurements. These measurements are then transmitted via the \ac{IoT} \textcolor{black}{network} to undergo remote cloud processing. \textcolor{black}{Despite the imposed single-epoch framework, the processing} power is not actually limited.}


\textcolor{black}{In this context,} the signals received from satellites in harsh environments such as urban canyons can be significantly impacted by \ac{NLOS} and \ac{MP} propagation, resulting in strongly biased pseudorange measurements. It is hence crucial to identify the most reliable and informative measurements while disregarding the most harmful ones. This 
 selection process is especially challenging as modern \ac{GNSS} receivers typically receive tens of measurements from different constellations at each time epoch. Basic selection approaches primarily rely on single-link satellite features, such as the signal carrier-to-noise power density ratio ($C/N_0$) \textcolor{black}{or} the elevation angle ($\theta$), to exclude or mitigate the impact of satellites that are likely to contribute to large positioning errors~\cite{CN0SigmaE,CN0SigmaDelta}. 

Several techniques have been proposed to select the remaining satellites that meet basic single-link quality criteria, including subset-testing \cite{subsetTesting}, \textcolor{black}{RANdom SAmple Consensus (RANSAC)} \cite{RANSAC}, iterative reweighting \cite{iterativeReWeighting}, etc. These methods offer various trade-off levels between computational complexity and performance. However, because of the massive combinatorial complexity of testing all possible subsets of satellites, exhaustive search is not feasible and the selection problem remains an open issue to the best of our knowledge. Conventional selection approaches \cite{RAIMcon}, rely on the spatial distribution of intermediate positioning results conditioned on specific satellite subsets to determine the most harmful contributions. 
These approaches only exclude satellites with strongly biased pseudoranges \textcolor{black}{instead of mitigating for these biases.} This may limit the best possible positioning accuracy accordingly.

In this paper, we present a new pre-processing approach specifically designed for single-epoch stand-alone positioning, which addresses the limitations of conventional satellites selection techniques. Our approach casts the initial satellite selection problem into a weighting problem. A first innovative aspect lies in exploiting \ac{ML} into the domain of pseudorange residuals, resulting from offline conditional positioning results, jointly with that of instantaneous SV measurements, such as $C/N_0$ and angle elevation, to optimize satellite weights.

\textcolor{black}{To fully exploit the potential of deep learning tools in uncovering the hidden correlations between pseudorange measurements and the position solution}, as well as possible joint effects of discarding multiple measurements at a time, we exploit a \acf{LSTM NN} from the family of \textcolor{black}{\acfp{RNN}} that is fed with a customized feature matrix, what represents another originality of our contribution. The LSTM NN predicts quality factors that estimate the link-wise standard deviations of pseudorange errors. These predictions are then used to compute nearly-optimal satellite weights within a conventional \ac{WLS} positioning solver. 

The training and testing of our \ac{NN} were performed using real data collected from extensive field tests comprising more than $290$ experiments \textcolor{black}{(i.e., driving sessions)} and $440,000$ epochs. \textcolor{black}{For this purpose, we used a receiver capable of operating over mulitple constellations in both single-band and dual-band modes.} 
The ground-truth reference positioning data was collected from a high-end GNSS-aided INS system providing cm-level accuracy.

In summary, the main contributions of this paper 
are: First, we propose a novel deep learning-based solution to the satellite selection problem. Our approach relies on a \ac{LSTM NN} fed with a \textcolor{black}{customized matrix containing joint features (i.e., conditional pseudorange residuals) and per-link features (e.g., carrier-to-noise power density ratio and its statistics, elevation angle, carrier phase lock time).} 
Second, our approach is shown to enhance the computation of measurement weights compared to conventional parametric methods. This has been verified through real-field data obtained by measurement campaigns and a dedicated test platform in various driving scenarios and environments.

\vspace{-0.1in}
\section{Problem Formulation}\label{sec:problem_formulation}

\vspace{-0.05in}
\subsection{Single Epoch Positioning}
\label{sec:single_epoch}
To simplify the presentation and notations, with no loss of generality regarding the proposed approach, we consider here a set of $N$ single band, single constellation, pseudorange  measurements $\{\rho^{i}\}_{i=1\hdots N}$. 
While accounting for experimental results in Section~\ref{sec:results}, we will consider a multi-band and multi-constellation scenario. The necessary corrections, computed from ephemeris data (such as \ac{SV} clock bias, ionospheric and tropospheric delays, Sagnac correction, etc.), have already been applied to these pseudorange measurements. The measurement from the $i$-th \ac{SV}, in the absence of \ac{MP} or any strong bias, can be modeled as follows:
\begin{equation}
    \rho^{i} = \sqrt{(x-x^i)^2+(y-y^i)^2+(z-z^i)^2} + c\ \delta + \eta^i,
\end{equation}
where $\rho^{i}$ is the pseudorange between the receiver and the \mbox{$i$-th} $\ac{SV}$, with $(x^{i},y^{i},z^{i})$ and $(x,y,z)$ the coordinates of the \mbox{$i$-th} satellite and the receiver, respectively. The parameter $c$ is the speed of light, and $\delta$ is the clock bias between the receiver and the considered constellation. $\eta^i$ is the observation noise, which represents both the receiver noise and errors resulting from troposphere and ionosphere extra delays, etc. Although ionosphere and troposphere residual errors (i.e., after correction from navigation message) are highly correlated over time, they are usually considered as independent and zero mean in single-epoch processing. Hence, we simply assume that the observation noise follows a centered Gaussian distribution $\eta^i \sim \mathcal{N}(0,\,(\sigma^{2})^i)$.

Our aim is to estimate the vector $\ve{X}=[x,y,z,\delta]^\top$ from the set of measurements $\{\rho^{i}\}_{i=1 \hdots N}$. A Solution that is efficient and widely used is then provided by the maximum-likelihood estimator (MLE) \cite{Rossi2018}, which simplifies to a \textcolor{black}{\ac{WLS} estimator} for our Gaussian noise model:
\vspace{-0.08in}
\begin{equation}
 \label{eq:MLE_Result}
 \hat{\ve{X}}= \arg \min_{\ve{X}} \ \sum_{i=1}^{N}\omega^{i}(\rho^{i}-h^{i}(\ve{X}) \big)^2,
 \end{equation}
where the observation function for the $i$-th satellite is defined as 
\vspace{-2mm}
\begin{equation}
 \label{eq:Observation_function}
 h^{i}(\ve{X}) = \sqrt{(x-x^i)^2+(y-y^i)^2+(z-z^i)^2} + c \ \delta,
 \end{equation}
and the \textcolor{black}{optimal weights} are equal to 
\begin{equation}
 \label{eq:WeightsDefinition}
 \omega^{i}=\frac{1}{(\sigma^2)^i}.
 \end{equation}
The solution can be computed using an optimization algorithm such as Levenberg-Marquardt or Gauss-Newton \cite{Levenberg-Marquardt_Algo}.

\subsection{GNSS Satellite Selection and Weighting Problems}

Generally, {an initial rough} \ac{SV} selection based on satellite elevation or $C/N_0$ thresholds is performed to exclude presumably strongly biased measurements.
Then, the standard deviation of the remaining measurements is estimated using an empirical function, for instance as follows~\cite{groves2013principles}:
\begin{equation}
\label{eq:SoTA_NoiseDef}
(\sigma^2)^{i}=\frac{1}{\sin^2{(\theta^i)}}\left(\sigma^2_{\rho Z}+\frac{\sigma^2_{\rho c}}{(C/N_0)^i}+\sigma^2_{\rho a} (a^2)^i\right),
\end{equation}
This function mainly depends on satellite elevation $\theta$, $C/N_0$, acceleration $a$, and other empirically calibrated coefficients ($\sigma^2_{\rho Z}, \sigma^2_{\rho c}, \sigma^2_{\rho a}$) that are hard to be fine-tuned. 

 However, some measurements could be strongly biased by \ac{MP} for instance, and violate the expected Gaussian model, resulting in a significant degradation of the solution accuracy. It is thus of primary importance to exclude these measurements from the solution, either by discarding them or by assigning them a zero weight, which is also called de-weighting. Such measurements can be efficiently detected at the navigation processor stage based on innovation monitoring tests for instance \cite{EKFInnovation}, but this requires the tracking filter to \textcolor{black}{have already converged} and the predicted state (i.e., position, receiver clock offset, etc.) to be accurate enough.  
 
For single epoch processing, no predicted solution is available and \ac{SV} selection relies on measurements only with very limited prior knowledge. This implies that the detection of $k$ faults among $N$ measurements could potentially result in a huge number of subsets, $C_k^{N}$, to test in case of an exhaustive search, which is intractable in real-time and even for post-processing. As an example, assuming at most $10$ faults among $40$ measurements would result in more than $847\times10^6$ subsets to test, which is computationally prohibitive.

\vspace{-1.5mm}
\subsection{Existing Works}\label{subsec:Existing_work}


In the field of \ac{GNSS}, improving the integrity of navigation solutions requires the identification and rejection of corrupt signals. \ac{FDE} techniques, such as classical \ac{FDE} \cite{knight2009comparison}, \cite{kuusniemi2007user}, advanced receiver autonomous integrity monitoring (ARAIM) \cite{zhai2018fault}, \cite{joerger2014solution}, brute force subset testing \cite{angrisano2013gnss}, and range consensus (RANCO) \cite{RAIMcon}, have been proposed in the literature. Recently, a new \ac{FDE} algorithm \cite{VTCPaper2021} has been put forward that utilizes both a standalone \ac{FDE} block based on a residual test using \ac{WLS} and an \ac{FDE}-based Extended Kalman Filter (EKF). The algorithm alternates between the two branches depending on a covariance matrix threshold. When the covariance matrix falls below a pre-defined threshold, the EKF utilizing \ac{FDE} is employed; otherwise, the \ac{FDE} is used on its own. This algorithm has shown significant improvements in accuracy compared to conventional state-of-the-art \ac{FDE} algorithms, and is hence used as a reference for benchmark purposes in this paper. However, since the focus is on single-epoch localization applications, only the standalone \ac{FDE} block is utilized in this study.

\vspace{-0.09in}
\section{Proposed System Architecture}\label{sec:proposed_system_architecture}
As previously mentioned, we focus on single-epoch standalone positioning based on pseudoranges without differential corrections, where measurements have just been pre-processed (see Section \ref{sec:single_epoch}).
Our main objective is to exploit the complex inter-dependencies and joint effects across multiple links using supervised deep learning. This will allow us to efficiently weight the contributions from all satellites.

\textcolor{black}{For this, we first need to build a comprehensive set of information inputs to our \ac{NN}. One novelty of our proposal is to concatenate both per-link and joint features, which will be detailed below, and feed them} as one single customized matrix into the \ac{RNN} (i.e., \ac{LSTM NN}). The network is then trained to predict the weights $\hat{\omega}^i$ related to the underlying distribution of pseudorange errors according to (\ref{eq:WeightsDefinition}). We expect our network to predict nearly null weights $\hat{\omega}^i \approx 0 $ for strongly biased satellites to be excluded.

\begin{figure*}
\resizebox{1\textwidth}{!}{\includegraphics{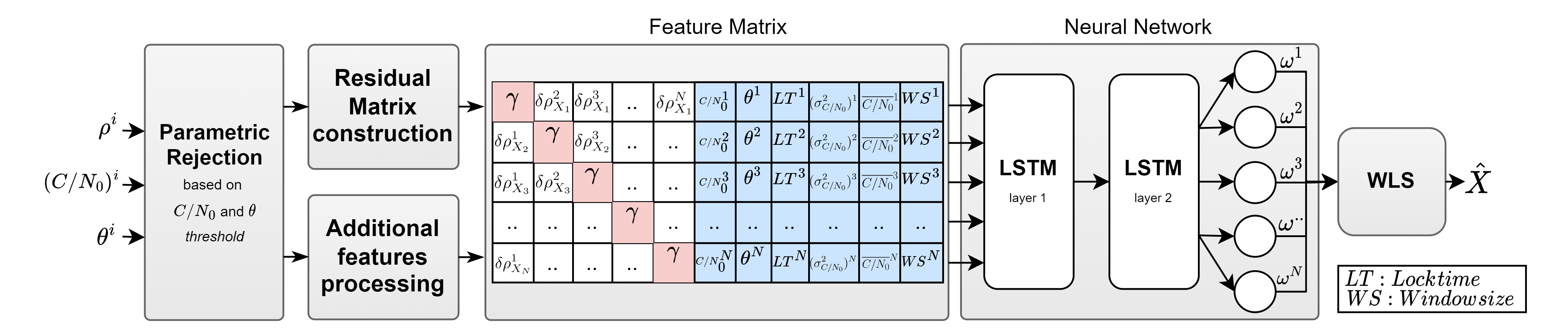}}
\vspace{-6mm}
\caption{Complete architecture of our proposed approach.}
\label{fig_ApproachDiagram}
\vspace{-6mm}
\end{figure*}

To sum up, the original difficult selection problem is cast into a soft weighting problem. Fig.~\ref{fig_ApproachDiagram} shows the complete architecture of our proposed solution.

\vspace{-2mm}
\subsection{Per-link Features}\label{Subsec:Per_Link_Features}

{\color{black}Several per-link \textcolor{black}{measurement} features can be informative, and hence beneficial to be provided as an input to the \ac{NN}, such as: the \ac{SV} elevation angle $\theta$, the carrier phase lock time, $C/N_0$ and its \textcolor{black}{empirical} statistics (i.e., empirical variance $\sigma_{C/N_0}^2$ and mean $\overline{C/N_0}$ computed for the sliding window).
 
 The intuition behind the \ac{SV} elevation angle as a feature is based on the fact that signals from \ac{SV} with low elevation angles have a longer traveling distance in the ionospheric and tropospheric layers and a greater probability of \ac{NLOS} conditions. Also, the carrier phase lock time is an important indicator for newly acquired \acp{SV}.
 
 Regarding $C/N_0$, low values increase the \textcolor{black}{tracking noise of the ranging processor} and, consequently, increase the pseudorange measurement noise. Beyond, the statistics of $C/N_0$ (i.e., its variance and mean) are also expected to convey informative indications about the operating conditions. 
For instance, it has been observed that the \textcolor{black}{short-term empirical} \textcolor{black}{variance}, computed over a short period of time, could be a fairly good indicator of \ac{MP} \cite{CN0Variablity}. 
 Even though we deal with single-epoch processing in our case, we can still practically approximate the actual \textcolor{black}{variance} of $C/N_0$  over a few past seconds, as it will only affect measurement weights and introduce minimal correlation between them.}

\begin{figure}
\resizebox{0.5\textwidth}{!}{\includegraphics{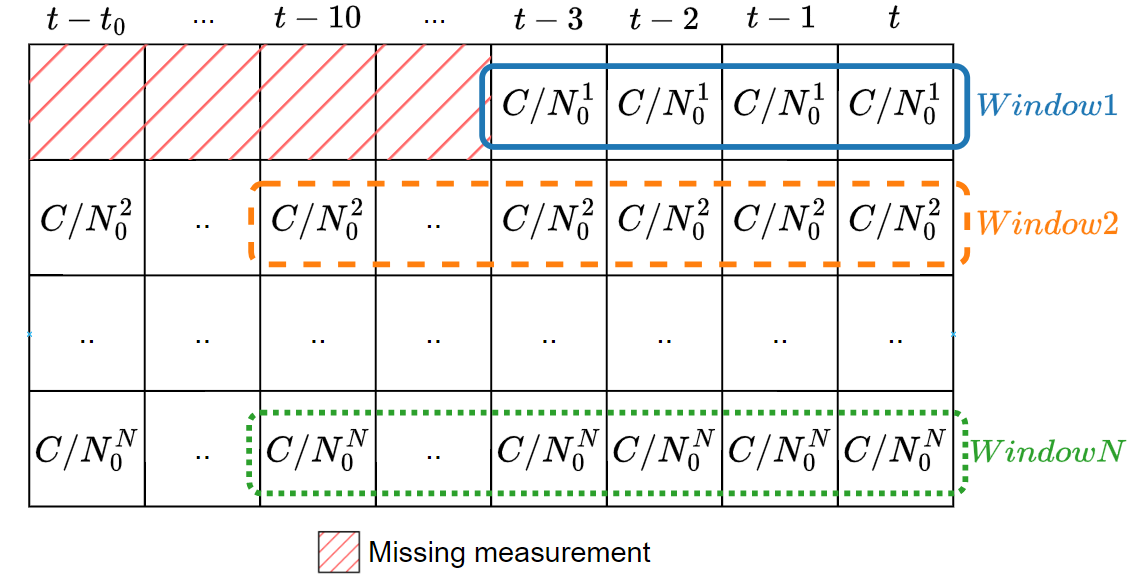}}
\vspace{-6mm}
\caption{Variable size sliding window for $C/N_0$ variance and mean calculation.}
\label{fig_SlidingWindow}
\vspace{-0.2in}
\end{figure}

\textcolor{black}{For this purpose, we consider a sliding window to calculate both the \textcolor{black}{empirical} variance $\sigma_{C/N_0}^2$ and mean $\overline{(C/N_0)^i}$ of $C/N_0$. This time window varies in size and contains the $C/N_0$ values from the previous epochs. As illustrated in Fig. \ref{fig_SlidingWindow}, the size of the window is determined by the number of epochs in which a particular satellite was received, and it can reach a maximum of $10$ epochs (equivalent to $2$ seconds). If the window contains only one entry, the variance is set to an arbitrarily large value. 
At last, the size of the sliding window is also fed as a feature to the \ac{NN}.}

As a result, the obtained per-link features are concatenated into a sub-matrix containing $6$ feature column vectors of dimension $N$ (see Fig.~\ref{fig_ApproachDiagram}).

\vspace{-0.09in}
\subsection{Joint Features}\label{subsec:Residual_matrices_construction}

Joint features \textcolor{black}{accounting for the simultaneous impact of multiple measurements over distinct links (i.e., in contrast to per-link features)} can also be extracted from the comparison of the \textcolor{black}{positioning} solutions from different tested subsets.  
To overcome the challenge of testing all the possible subset combinations, which is computationally prohibited, our approach leverages the ability of \ac{ML} in extracting hidden inter-dependencies from only $N$ such subsets. 

At each navigation epoch, we assume that multiple (i.e., $N$) satellite signals are received and a new matrix of positioning residuals \textcolor{black}{$\mathbf{M}$} is constructed, as follows. We generate $N$ subsets \textcolor{black}{$S_n$} of $N-1$ satellites, where we exclude one distinct satellite \textcolor{black}{(i.e, $n^{th}$ satellite)} at a time.
\begin{equation}
\label{eq:SubsetDef1}
S_n=\{\rho^i\},~i=1 \hdots N,~i\neq n
\vspace{-0.08in}
\end{equation}

For each subset $S_n$, we calculate the corresponding solution $X_{n}$  using (\ref{eq:MLE_Result}) with equal weights. Then, for each of the $N$ resulting positions $\{\ve{X}_{1},..., \ve{X}_{N}\}$ we calculate the $N-1$ pseudorange residuals 
\begin{equation}
\label{eq:SubsetDef2}
\delta\rho_{{X}_n}^i=\rho^i-h^i(\ve{X}_n),~ ~i\neq n
\vspace{-1mm}
\end{equation}
The coefficient $[\mathbf{M}]_{n,i}$ (i.e. row $n$, column $i$) of the residual matrix $\mathbf{M}$ is then simply given by the corresponding residual for non-diagonal coefficients, or by an arbitrary large value $\gamma$ for the diagonal terms, indicating that the satellite has been deliberately excluded.
\begin{equation}
\label{eq:SubsetDef3}
[\mathbf{M}]_{n,i}=
\begin{cases}
      \delta\rho_{X_n}^i,~ ~i\neq n \\
      \gamma, ~i=n
\end{cases} 
\end{equation}
Each row $n$ of the matrix will thus provide residuals associated with the exclusion of the $n$-th measurement. Although it assumes a single fault per subset (i.e., row), our intuition is that such a matrix is able to \textcolor{black}{reveal the complex joint contributions of each satellite onto the positioning solution,} while being fed as a single input to the \ac{NN}.  
\vspace{-2mm}
\subsection{Long-Short Term Memory Neural Network}
\textcolor{black}{The overall input matrix of features being fed to the \ac{NN} \textcolor{black}{can be seen as} a sequence of $N$ pseudo-observations. At each observation, a single pseudorange measurement is excluded from computing the solution. By analyzing this sequence, the \ac{LSTM NN} can exploit the correlations between the excluded measurement and the solution and pinpoint which measurement exclusions have the best impact on the quality of the positioning solution. As a result, the \ac{NN} will be capable of predicting weights that exclude multiple biased measurements by analyzing the sequence of pseudo-observations, \textcolor{black}{set as joint features (see Section~\ref{subsec:Residual_matrices_construction}).} Besides, the additional per-link features for the excluded measurements were concatenated for each pseudo-observation to provide more information about the excluded satellites} \textcolor{black}{(see Section~\ref{Subsec:Per_Link_Features}).}

In this kind of problems, the \ac{LSTM NN} architecture, which is a type of \ac{RNN} \cite{huang2019deep}, has the advantage of keeping memory over multiple (possibly distant) pseudo time steps. Hence, \textcolor{black}{it} is also suited to exploit the correlations across the matrix rows in our case, even if we explicitly deal with a single-epoch problem. Similar applications of the \ac{LSTM NN} to other time-invariant problems have already been considered. For instance, in \cite{HandwritingRec}, \ac{LSTM NN} was used to process data with long-range interdependence (i.e., using geometric properties of the trajectory for unconstrained handwriting recognition). 

\textcolor{black}{Note that several other (more complex) \ac{NN} architectures were evaluated. For instance, we considered \textcolor{black}{a more complex architecture composed of two different concatenated \acp{NN}}. The first \ac{NN} processes only the residual matrix as input. Its output is concatenated with the additional per-link features and fed as inputs to the second fully-connected NN (FCNN). However, such architectures did not provide any significant improvement over the LSTM NN that only processes the residual matrix. For the sake of conciseness, such alternative architectures are not further discussed in this paper.}

\vspace{-1mm}
\section{Numerical Results}\label{sec:results}
\textcolor{black}{To assess both the feasibility of our approach and the impact of using a customized residual matrix to capture inter-dependencies between the measurements and the computed solution, we conducted two tests. First, we evaluated our approach using only the residual matrix as input to the \ac{NN}. Second, we evaluated our approach using the overall feature matrix \textcolor{black}{(containing the residuals coupled with the per-link features)} as an input. The results obtained were compared with the \ac{SOTA} approach~\cite{VTCPaper2021} described in Section~\ref{subsec:Existing_work}.}
\vspace{-1mm}
\subsection{Data Collection and Scenarios}
\textcolor{black}{We conducted our tests on real-world data collected from measurement campaigns under various operational conditions. These conditions included open skies, dense urban areas, and a range of mobility scenarios. The data was collected using Vehloc, a multi-sensor platform developed by CEA-Leti. This board can be equipped \textcolor{black}{with one out of two different receiver chips:} either a NEO-M8P or a ZED-F9P, both from Ublox. The ZED-F9P dualband RTK GNSS \textcolor{black}{chip} can receive up to $N=60$ satellite signals from various \ac{GNSS} constellations (including GALILEO, GPS, GLONASS, etc.) across multiple frequencies (i.e., L1, L2, E1, and E5 bands). The NEO-M8P \textcolor{black}{chip} can receive single-band L1 signals from multi-constellation (i.e., \textcolor{black}{only GPS and GLONASS were used}).} 

\begin{figure}
\center
\resizebox{0.4\textwidth}{!}{\includegraphics{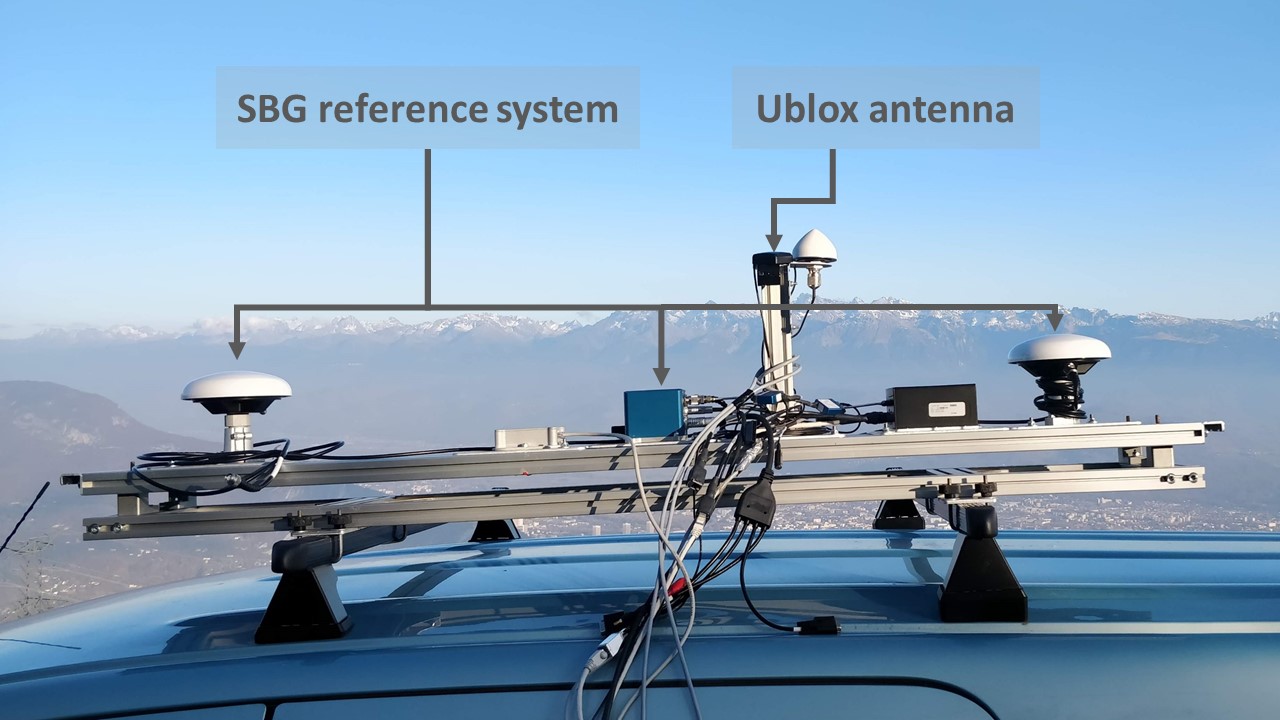}}
\vspace{-0.1in}
\caption{Data collection setup.}
\label{fig_Setup}
\vspace{-0.2in}
\end{figure}

The raw measurements were collected at a rate of $5$ Hz, as well as the broadcasted ephemeris. These measurements included the \ac{SV} ID, constellation ID, frequency ID, pseudorange, pseudorange rate, carrier phase, $C/N_0$, carrier phase lock time, etc. 
The reference position was obtained from a high-end, GNSS-aided INS,  Ekinox platform from SBG, which was post-processed using Qinertia software (from SBG) to provide centimeter-level accuracy, even in difficult GNSS conditions. The complete experimental setup is depicted in Fig. \ref{fig_Setup}, which shows the Ublox antenna of the Vehloc platform and the SBG reference system mounted on the roof of a vehicle.
In total, $291$ experiments were conducted during the data collection campaigns phase, which yielded a diverse, representative, and comprehensive dataset for evaluation, encompassing up to $440,000$ epochs.

\vspace{-2mm}
\subsection{Evaluation Results}
\textcolor{black}{The architecture of our \ac{NN} including the number of layers and the number of neurons per layer was optimized empirically. So as to prevent overfitting effects, an early stopping callback was used during the training process. Data was randomly split into three disjoint subsets: 60\% of the data for training, 20\% of the data for validation, and 20\% for testing and evaluating the performance of our approach (i.e., unseen during training phase).}
The best architecture among the tested ones was then found to consist of $2$ hidden layers composed of $893$ neurons each (see Fig.~\ref{fig_ApproachDiagram}).  

\begin{figure}
\resizebox{0.5\textwidth}{!}{\includegraphics{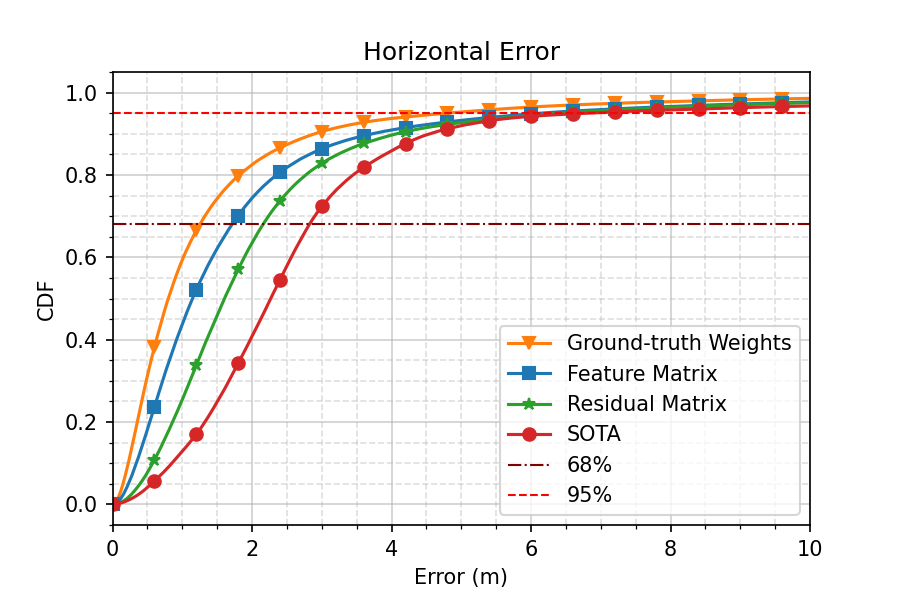}}
\vspace{-8mm}
\caption{Empirical \ac{CDF} of horizontal error for various measurements weighting/selection strategies (incl. \cite{VTCPaper2021}) on real field data \textcolor{black}{(over all navigation sessions from ZED-F9P receiver)}.}
\label{fig_RealData_HorError}
\vspace{-0.2in}
\end{figure}
\begin{figure}
\vspace{-2mm}
\resizebox{0.5\textwidth}{!}{\includegraphics{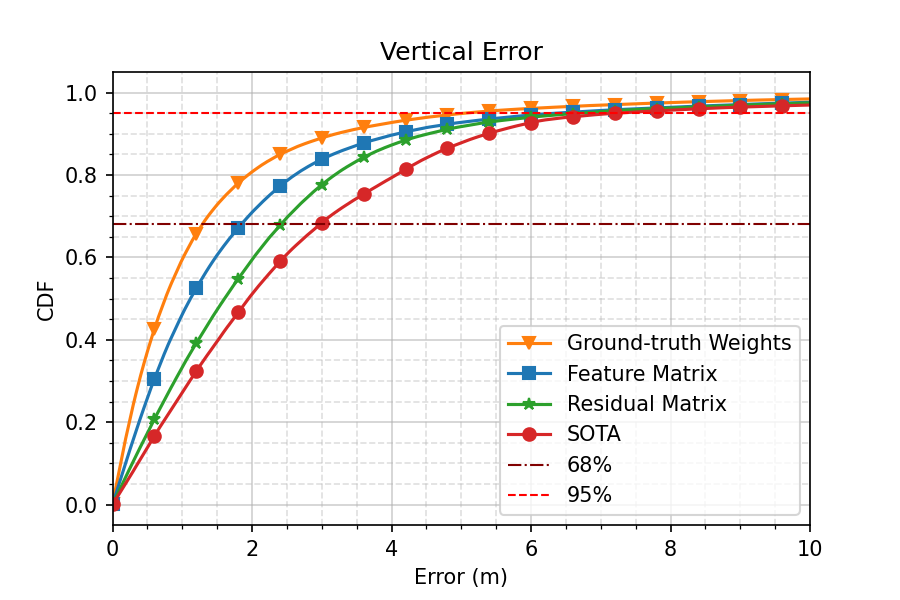}}
\vspace{-8mm}
\caption{Empirical \ac{CDF} of horizontal error for various measurements weighting/selection strategies (incl. \cite{VTCPaper2021}) on real field data \textcolor{black}{(over all navigation sessions from ZED-F9P receiver)}.}
\label{fig_RealData_VerError}
\vspace{-0.2in}
\end{figure}

In supervised learning, the data must be labeled. However, the optimal labels are not well-defined \textcolor{black}{in our case}. From the Bayesian estimation perspective, this would require having access to the true distribution (i.e., the standard deviation in particular, assuming a Gaussian distribution of the measurements errors), while only a single sample of this distribution is available. Indeed, having more precise information about the underlying distribution would require collecting multiple measurements from the same receiver's position with also the same \ac{SV} positions, 
 \textcolor{black}{which is not feasible}. When dealing with a single sample/observation, the most accurate estimate of the standard deviation is \textcolor{black}{the absolute value of} the error calculated specifically for that measurement. This approach is further reinforced by empirical evidence, by utilizing the weights as:
\vspace{-0.1in}
\begin{equation}\label{eq:Weights_truth}
    \textcolor{black}{\omega^{i}=1/(\rho^{i}-h(\ve{X}_{true}))^2},
\vspace{-1.5mm}
\end{equation}
where the $\ve{X}_{true}$ stands for the ground-truth position collected from the reference system. Utilizing this weighting method has provided very good results as shown in Fig. \ref{fig_RealData_HorError} and \ref{fig_RealData_VerError}, thus these weights were used as the data labels.

\textcolor{black}{As a first example, we have analyzed a vehicular setup of $40$ minutes, comprising urban, sub-urban and back-country environments \textcolor{black}{using the ZED-F9P receiver chip}.}
For this test, the horizontal error for the \ac{SOTA} single-epoch algorithm is only $2.18$ m, which can be qualified as good when compared to the accuracy of the $2.03$ m solution provided by the receiver benefiting from a filtered mode. Our approach offers a significant performance improvement (of $35\%$) with only $1.41$ m of error.

\begin{figure}
\centering
\resizebox{0.4\textwidth}{!}{\includegraphics{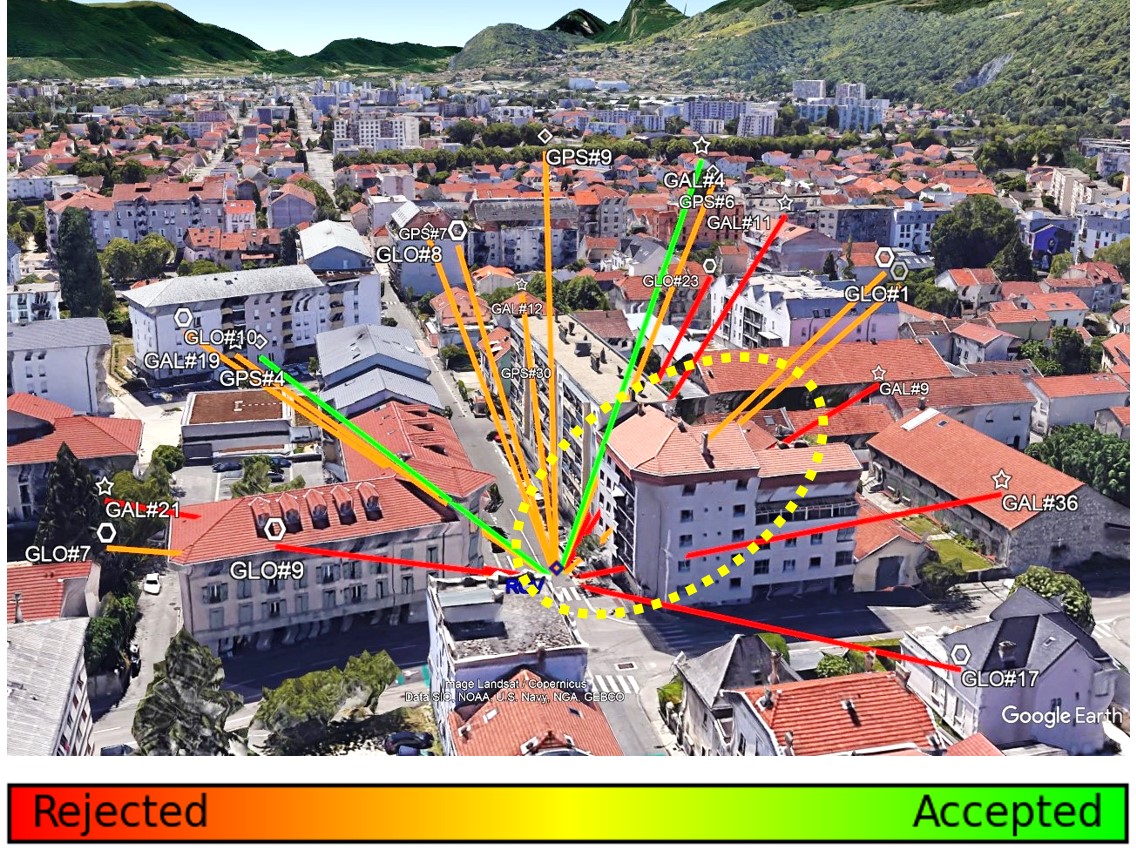}}
\vspace{-0.1in}
\caption{\textcolor{black}{Conceptual illustration of satellite measurements weighting with the proposed approach.}}
\label{fig_3DSat_Ourapproach}
\vspace{-0.15in}
\end{figure}

\begin{figure}
\centering
\resizebox{0.4\textwidth}{!}{\includegraphics{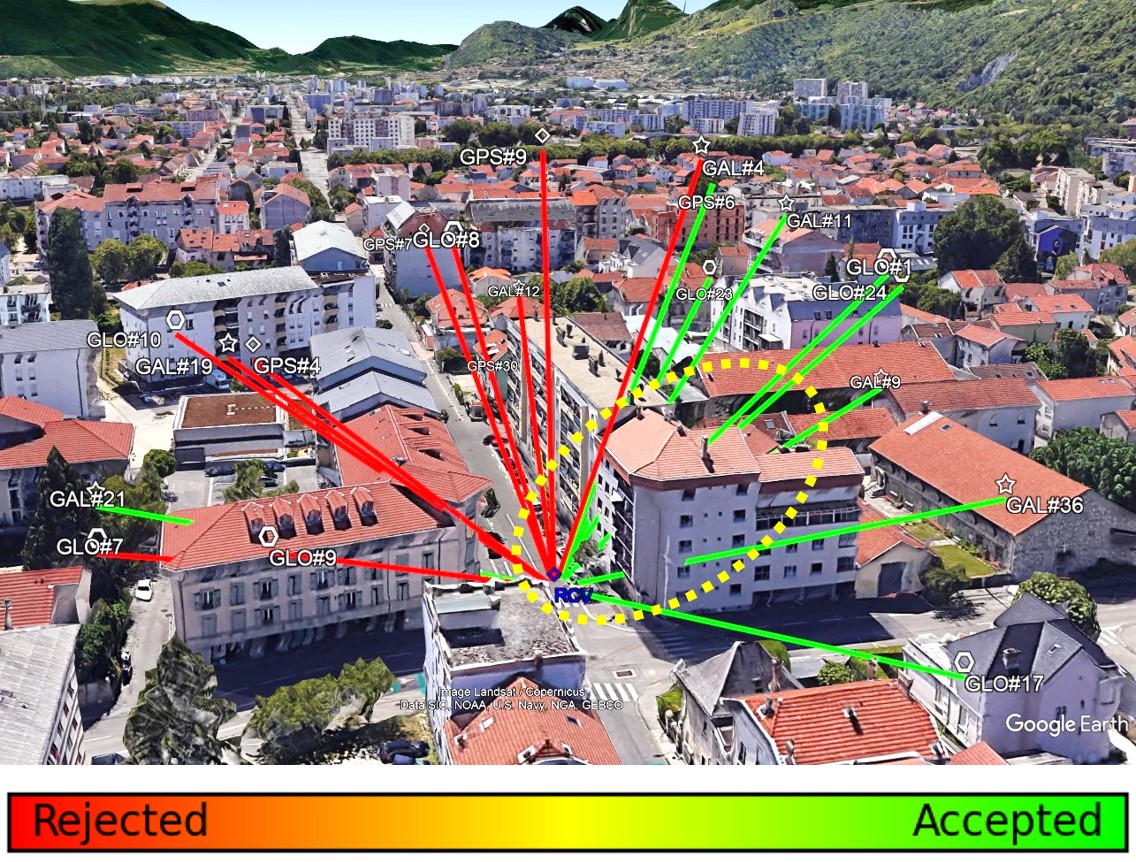}}
\vspace{-0.1in}
\caption{\textcolor{black}{Conceptual illustration of satellite measurements weighting with the \ac{SOTA} approach~\cite{VTCPaper2021}, for the same epoch as that of Fig.~\ref{fig_3DSat_Ourapproach}.}}
\label{fig_3DSat_SOTA}
\vspace{-0.2in}
\end{figure}

\textcolor{black}{To provide a better understanding of the possible benefits of our approach, we \textcolor{black}{focus next} on a specific epoch in a sub-urban environment, where the difference with \ac{SOTA} is quite significant.}
\textcolor{black}{For qualitative benchmark purposes,} Fig. \ref{fig_3DSat_Ourapproach} and \ref{fig_3DSat_SOTA} illustrate the satellite links in this setup and how the satellites were treated \textcolor{black}{by the two approaches}, where the color bar indicates the degree (between accepted or rejected) of the weights used for positioning. As it can be seen, \acp{SV} GAL\#36, GAL\#9, GLO\#24, GLO\#1, etc. are clearly in \ac{NLOS} conditions. These \acp{SV} were not rejected by the \ac{SOTA} approach (see Fig. \ref{fig_3DSat_SOTA}), while they were either rejected or weighted by our approach (see Fig. \ref{fig_3DSat_Ourapproach}). Furthermore, other \acp{SV} in LOS conditions such as  GPS\#9, GPS\#7, GLO\#8, GAL\#12, etc. were completely rejected by the \ac{SOTA} approach, while being efficiently weighted by our approach.

\begin{figure}
\resizebox{0.5\textwidth}{!}{\includegraphics{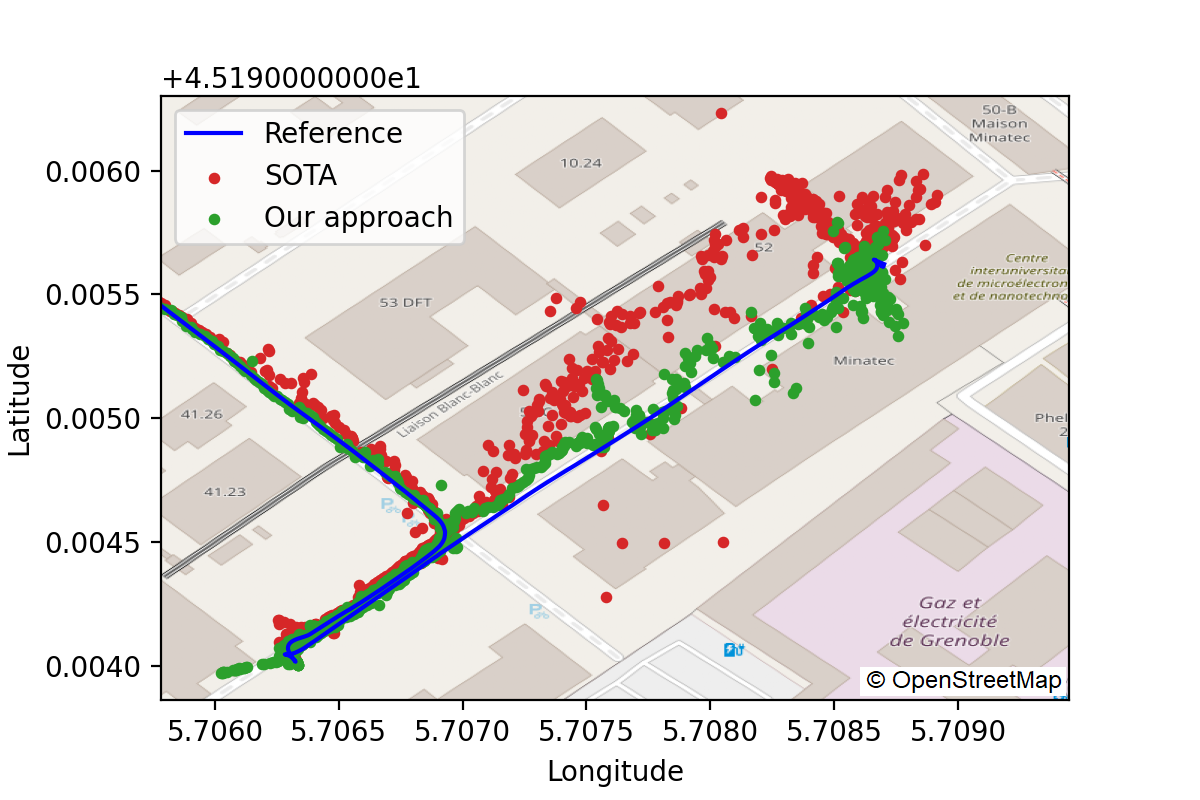}}
\vspace{-8mm}
\caption{Example of positioning traces obtained in a field navigation session (urban canyon) with the proposed approach (green dots), the one in \cite{VTCPaper2021} from recent state-of-the-art (red dots), and the ground-truth reference system (blue solid line).}
\label{fig_MapDist}
\vspace{-0.2in}
\end{figure}

\textcolor{black}{Fig. \ref{fig_MapDist} represents another comparison of our approach with \ac{SOTA}, 
focusing on \textcolor{black}{a specific portion of one navigation session in} a GNSS-challenging environment (urban canyon). The deviation of the localized position with our approach from the reference one is much less than that of the \ac{SOTA} approach. \textcolor{black}{In this case, several strongly biased \acp{SV} in \ac{NLOS} conditions, such as GPS\#12, GPS\#32, and GLO\#9, were used by the \ac{SOTA} approach to calculate a position estimation, whereas in our approach the same \acp{SV} have been excluded from the solution by assigning them very low weights (i.e., $\omega^{GPS\#12} = 0.0016, \ \omega^{GPS\#32} = 0.0006, \ \omega^{GLO\#9} = 0.002$), hence resulting in a much better accuracy of $2.06$ m typically}. In other words, our approach \textcolor{black}{seems more robust and suited to highly challenging environments, by mitigating the influence of strongly biased measurements through adequate weighting.}} 

Finally, Fig.~\ref{fig_RealData_HorError} and \ref{fig_RealData_VerError} show the empirical \ac{CDF} of horizontal and vertical positioning errors obtained with four different weighting strategies \textcolor{black}{over all navigation sessions from ZED-F9P receiver (386,000 epochs)}: the first one, referred to as ``Ground-truth Weights'', utilizes weights as in (\ref{eq:Weights_truth}) with the ground-truth position, before applying \ac{WLS} positioning. 
The second method, named ``Feature Matrix'', exploits
\textcolor{black}{the overall feature matrix comprising the two sub-matrices (i.e., the ``Residual Matrix'' concatenated with the additional per-link features)} as an input to the \ac{NN} to predict link-wise quality factors which are then used to compute weights as in  (\ref{eq:WeightsDefinition}) for \ac{WLS} positioning.
The third method, named ``Residual Matrix'', uses \textcolor{black}{only the residual matrix (joint features) as an input to the proposed algorithm.} 
Finally, in the fourth ``SOTA'' method, the state-of-the-art solution from \cite{VTCPaper2021} (See Section~\ref{subsec:Existing_work}) is used for positioning. 

We thus observe a significant improvement in terms of both horizontal and vertical errors of the deep learning approaches compared to the state-of-the-art solution. On the one hand, ``Feature Matrix'' exhibits a performance gain with respect to the \ac{SOTA} approach of $1$ m (resp. $1.47$ m) in terms of horizontal error at $68$\% (resp. $95$\%) of the \ac{CDF}. As for the vertical error, we also observe an improvement of $1.2$ m (resp. $1.75$ m) at $68$\% (resp. $95$\%) of the \ac{CDF}. On the other hand, the ``Residual Matrix'' approach shows a performance gain compared to the \ac{SOTA} approach of $0.61$ m (resp. $1.38$ m) in terms of horizontal error at $68$\% (resp. $95$\%) of the \ac{CDF}. As for the vertical error, we also observe an improvement of $0.66$ m (resp. $1.43$ m) at $68$\% (resp. $95$\%) of the \ac{CDF}. 

It is worth highlighting that the additional per-link features concatenated with the residuals have a significant and positive impact on the positioning accuracy. The superiority of the proposed approaches on the tested data illustrates their relative robustness in various operating conditions.  

\vspace{-1mm}
\section{Conclusions}
\vspace{-1mm}

\textcolor{black}{In this paper, we introduced a novel approach for \ac{SV} measurement weighting in single-epoch \ac{GNSS} positioning by exploiting a combination of joint features (i.e., conditional pseudorange residuals) and per-link features (i.e., $C/N_0$ and its empirical statistics, \ac{SV} elevation angle, carrier phase lock time). Our approach relies on a \ac{LSTM NN} in predicting several quality factors to weight the contributions of different measurement into the calculation of a position. Our results on real data obtained from field experiments demonstrated the robust performance of the proposed solution in challenging environments, while outperforming a recent \ac{SOTA} approach. This solution is particularly promising into IoT applications (remote processing), where accurate single-epoch positioning is essential, whether in real-time or offline.}




\section*{Acknowledgement}
The work included in this paper has been partly supported by the “France 2030” investment program through the project 5G-OPERA.

\color{black}
\bibliographystyle{IEEEtran}
\bibliography{main}

\vspace{12pt}

\end{document}

%% file: acronyms.tex
\usepackage[nolist]{acronym}

\newacro{IoT}{Internet of Things}
\newacro{INS}{Inertial Navigation System}
\newacro{GNSS}{Global Navigation Satellite System}
\newacro{LSTM NN}{long-short term memory  neural network}
\newacro{LSTM}{long-short term memory}
\newacro{NN}{neural network}
\newacro{ML}{machine learning}
\newacro{WLS}{weighted least squares}
\newacro{FDE}{fault detection and exclusion}
\newacro{RNN}{recurrent neural network}
\newacro{MSE}{mean squared error}
\newacro{CRLB}{Cramer Rao lower bound}
\newacro{CDF}{cumulative density function}
\newacro{SV}{satellite vehicle}
\newacro{SOTA}{state-of-the-art}
\newacro{NLOS}{Non-Line of Sight}
\newacro{MP}{multi-path}